# Tell Me the Good Stuff: User Preferences in Movie Recommendation Explanations

Juan Ahmad*
jwan.k.ahmed@gmail.com
University of Gothenburg
Gothenburg, Sweden

Jonas Hellgren*
hellgren.jonas@gmail.com
University of Gothenburg
Gothenburg, Sweden

Alan Said
alan@gu.se
University of Gothenburg
Gothenburg, Sweden

## ABSTRACT

Recommender systems play a vital role in helping users discover content in streaming services, but their effectiveness depends on users understanding why items are recommended. In this study, explanations were based solely on item features rather than personalized data, simulating recommendation scenarios. We compared user perceptions of one-sided (purely positive) and two-sided (positive and negative) feature-based explanations for popular movie recommendations. Through an online study with 129 participants, we examined how explanation style affected perceived trust, transparency, effectiveness, and satisfaction. One-sided explanations consistently received higher ratings across all dimensions. Our findings suggest that in low-stakes entertainment domains such as popular movie recommendations, simpler positive explanations may be more effective. However, the results should be interpreted with caution due to potential confounding factors such as item familiarity and the placement of negative information in explanations. This work provides practical insights for explanation design in recommender interfaces and highlights the importance of context in shaping user preferences.

## 1 INTRODUCTION

Recommender systems have become essential tools for helping users navigate vast content libraries on streaming platforms. With catalogs containing thousands of titles, these systems assist users by filtering and suggesting content based on past behaviors and preferences [8, 9, 15]. However, these systems often appear as a "black box" to users [1], particularly when the reasoning behind suggestions is not clearly communicated. This study explores explanation design using non-personalized, feature-based item descriptions to simulate recommendations.

A key challenge in recommender systems is providing explanations that help users understand and evaluate suggestions. Research indicates that user trust and satisfaction can be influenced by the transparency of the system — that is, how well it conveys the reasoning behind recommendations [17, 18]. Previous work has explored a range of explanation styles, including collaborative filtering-based and feature-based formats that highlight specific item attributes [7, 20].

Recent work suggests that being transparent about both positive and negative aspects of recommendations — a two-sided approach — might increase user trust. Some studies in high-stakes domains have found that two-sided explanations can enhance trust and decision confidence [12, 13]. However, other work, especially in hedonic or low-stakes domains, suggests that one-sided messages can be more persuasive and credible [11]. This highlights a potential domain-dependence in how users respond to explanation styles.

This question is particularly relevant for streaming services, where content libraries continue to expand rapidly. Major platforms now offer thousands of titles across movies, TV shows, and other content types, making effective recommendation explanations increasingly important for user satisfaction and retention. While prior studies have evaluated explanation formats, few have examined how the balance of content (positive vs. mixed valence) affects user perceptions specifically in the domain of movie recommendations using natural-language, feature-based explanations. As the majority of explanations in recommender systems research are natural-language based [10], understanding how to design these explanations effectively is crucial.

Our work investigates how users perceive and respond to different styles of feature-based explanations in movie recommendations. Specifically, we address the following research questions:

**RQ1:** How do users perceive two-sided explanations compared to one-sided explanations in terms of transparency, trust, and effectiveness?

**RQ2:** Does the inclusion of negative features in explanations affect users' satisfaction with the recommender system?

**RQ3:** How do different aspects of explanation perception (transparency, trust, effectiveness) correlate with overall satisfaction?

To explore these questions, we conducted an online study with 129 participants, comparing user perceptions of one-sided and two-sided feature-based explanations in movie recommendation scenarios. Our results suggest that explanation design preferences may differ from established assumptions and are shaped by the domain and context of use. These findings offer preliminary insights for designing more user-aligned recommendation interfaces in entertainment settings.

## 2 RELATED WORK

Recommender systems help users identify relevant content by filtering large item catalogs typically using collaborative filtering or content-based approaches [6]. While algorithmic performance is essential, user-centered factors such as trust and transparency are equally important for success. A key challenge is the limited transparency of many systems, which often appear as "black boxes" to users [1]. Explanations aim to address this by clarifying why particular items are recommended. In their seminal work, Tintarev and Masthoff [18] outlined seven goals for explanations—transparency, scrutability, trust, effectiveness, persuasiveness, efficiency, and satisfaction—highlighting their interdependence. For instance, Sinha and Swearingen [17] demonstrated that transparency enhances both trust and satisfaction.

A growing body of work has explored explanation types, especially natural-language, feature-based explanations that describe specific item attributes [2, 20]. These explanations are often more informative than user-based or rating-based approaches and are increasingly important as recommender interfaces adopt conversational or textual formats [10].

One area of ongoing debate is whether explanations should include only positive aspects (one-sided) or a mix of positive and negative information (two-sided). Pu and Chen [13] and Pu and Chen [12] found that two-sided explanations can increase trust and confidence, particularly when trade-offs are expected. However, other work suggests this benefit may not generalize. For example, Pentina et al. [11] showed that one-sided user reviews were perceived as more trustworthy and credible. Similarly, Chen and Wang [3] found that for experience goods like movies, one-sided reviews were more helpful than balanced ones. These studies suggest that perceived usefulness of valenced content may depend on the domain and user expectations.

Zhang et al. [20] explored multiple variables in explanation design—such as feature count, modality, and valence—within restaurant recommendations. While they found no significant differences between one-sided and two-sided explanations in that context, their results were inconclusive for valence, and the setting was distinct from entertainment media. Our work builds on theirs by isolating valence (positive-only vs. mixed) within movie recommendations, using natural-language explanations and measuring both quantitative and qualitative user responses.

Recent studies have also explored the role of large language models (LLMs) in generating explanations. For example, Said [14] and Silva et al. [16] examined the use of ChatGPT for generating personalized recommendations and explanations. While LLMs offer flexibility and fluency, Silva et al. [16] found that users did not necessarily rate personalized explanations as more effective than generic ones. These findings highlight the importance of explanation content and framing over personalization alone.

Finally, domain-specific research has revealed varying preferences for explanation style. Muhammad et al. [9] found that two-sided explanations were well received in hotel recommendations, where decisions involve higher cost or risk. In contrast, in domains like movie or music recommendations—where decisions are more hedonic and low-stakes—simpler, positively framed explanations may better match user needs. Our study contributes to this discussion by focusing explicitly on the entertainment domain and examining how explanation valence affects user perceptions of trust, transparency, and satisfaction.

## 3 METHOD

Our study investigated user perceptions of one-sided versus two-sided explanations in movie recommendations through an online experiment. We recruited participants via online platforms (Surveyswap, Surveycircle) and social media channels. Participants were not compensated. After excluding 11 responses due to incompletion or failed attention checks, we analyzed data from 129 participants (45% male, 54.2% female, 0.8% identifying as other), with the majority (84.5%) between ages 18–34. This age skew may affect the generalizability of our results.

The experiment employed a repeated measures within-subject design, with explanation type (one-sided vs. two-sided) as the independent variable and participant perceptions of satisfaction, trust, transparency, and effectiveness as dependent variables. We implemented the survey using Psytoolkit, presenting participants with eight evaluation tasks divided into two blocks of four, separated by an attention check. Each block contained two one-sided and two two-sided explanations in randomized order. Participants were instructed to evaluate the style and clarity of the explanations, not the movie itself or its match to their own preferences.

Each task presented participants with a movie poster accompanied by an explanation for why the movie was recommended. Fig. 1a shows an example of a two-sided explanation for *Inglourious Basterds*, highlighting its "unique take on World War II and variety of popular actors" as positive features, while noting "it contains graphic violence" as a negative feature. Fig. 1b demonstrates a one-sided explanation for *Interstellar*, emphasizing only positive aspects: "breathtaking visuals, immersive tale, emotionally rich musical score."

The explanations were initially generated using ChatGPT and then manually refined to ensure clarity and consistency. To ensure consistency and control across conditions, we followed a semi-structured template during manual refinement of the ChatGPT-generated explanations. Each explanation was edited to include exactly three discrete features and written in a single-sentence format with parallel grammatical structure. For two-sided explanations, the negative feature was always introduced with a contrastive cue ("however") and positioned last. This placement decision was informed by a desire to mimic typical narrative phrasing, but we acknowledge it may have introduced recency effects. Positive features were selected based on commonly cited attributes in user reviews (e.g., acting, cinematography, soundtrack) and were chosen to be broadly appealing. Negative features were framed as mild drawbacks (e.g., violence, slow pacing) to avoid overtly polarizing the explanation. Two independent authors reviewed all explanations to ensure consistent tone, length, and linguistic quality across conditions.

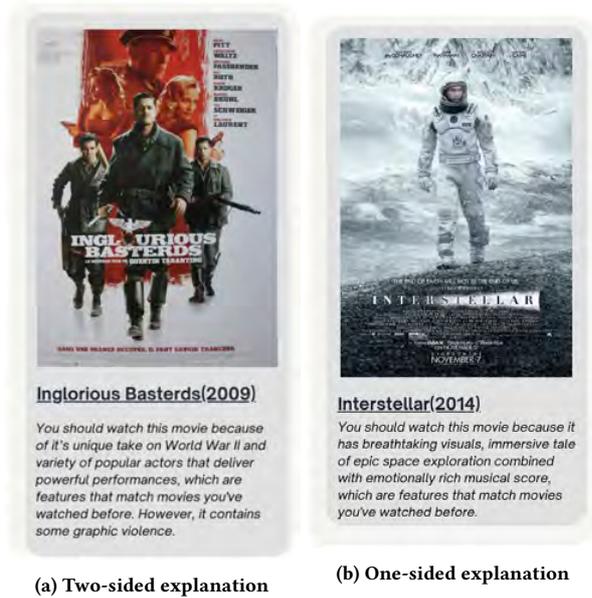

(a) Two-sided explanation  (b) One-sided explanation

Figure 1: Examples of explanation styles shown to participants. Fig. 1a Two-sided explanation for the film *Inglourious Basterds* (2009), highlighting two positive features and one negative feature. Fig. 1b One-sided explanation for the movie *Interstellar* (2014), featuring three positive features.

| | |
|---|---|
| Satisfaction: | I would enjoy using a recommender system if it presented recommendations in this way. |
| Trust: | I doubt that the system will recommend movies that I will like. |
| Transparency: | I understand why the item is recommended. |
| Effectiveness: | The information is sufficient for me to assess the recommended item. |

Table 1: Likert-scale statements used to assess participants' perceived satisfaction, trust, transparency, and effectiveness of explanations.

| Question | Type |
|---|---|
| Is it important for you to understand why a movie is recommended to you? | Yes/No |
| What kind of information do you think you need in an explanation to make use of it? | Open ended |
| Feel free to express any other thoughts about the survey or the explanations. | Open ended |

Table 2: Questions asked to study participants after completion of the explanation tasks, and the type of question.

Participants rated four statements (Table 1) after each task using 5-point Likert scales (1 = "Strongly Disagree", 5 = "Strongly Agree"). The statement used to assess trust was phrased negatively ("I doubt that the system will recommend movies that I will like") to reduce leading responses. For analysis, this item was reverse-coded so that higher values indicate higher trust. After completing all evaluation tasks, participants answered several closed and open-ended questions (Table 2) to provide further context and qualitative feedback.

We calculated mean ratings for each participant's perception of one-sided and two-sided explanations across the four measures. We used paired $t$-tests to compare the differences between conditions and calculated Cohen's $d$ for effect sizes. Although Likert data are ordinal, paired $t$-tests are robust for within-subject comparisons with near-normal distributions. To validate results, we also conducted Wilcoxon signed-rank tests. To examine relationships between perceptual measures, we calculated Pearson correlation coefficients between satisfaction and the other three dimensions for each explanation type.

## 4 RESULTS AND ANALYSIS

Analysis of participants' ratings revealed consistent differences between one-sided and two-sided explanations across all four measured dimensions. As shown in Fig. 2, one-sided explanations received significantly higher ratings for satisfaction ($M = 3.38$, $SD = 0.83$)[1] compared to two-sided explanations ($M = 3.02$, $SD = 0.89$); $t_{128} = 5.62$, $p < .05$. Trust was also rated higher for one-sided explanations ($M = 3.31$, $SD = 0.81$) than for two-sided ones ($M = 2.99$, $SD = 0.76$); $t_{128} = -5.49$, $p < .05$. Similarly, one-sided explanations were perceived as more transparent ($M = 3.80$, $SD = 0.69$ vs. $M = 3.48$, $SD = 0.72$; $t_{128} = 6.13$, $p < .05$) and more effective ($M = 3.44$, $SD = 0.83$ vs. $M = 3.18$, $SD = 0.82$; $t_{128} = 4.0$, $p < .05$). Cohen's $d$ values ranged from 0.353 to 0.540, indicating small to medium effect sizes. These findings suggest a consistent but moderate preference for one-sided explanations across all measured aspects.

To better understand how the four perceptual measures relate to one another, we analyzed Pearson correlation coefficients between satisfaction and the other three dimensions (trust, transparency, and effectiveness) for each explanation type. As shown in Table 3, the correlations were consistently stronger in the one-sided condition. For one-sided explanations, satisfaction was strongly correlated with trust ($r = .509$, $p < .05$), transparency ($r = .577$, $p < .05$), and effectiveness ($r = .697$, $p < .05$). For two-sided explanations, the correlations were moderate to strong: trust ($r = .475$, $p < .05$), transparency ($r = .443$, $p < .05$), and effectiveness ($r = .569$,

---

[1] mean and standard deviation

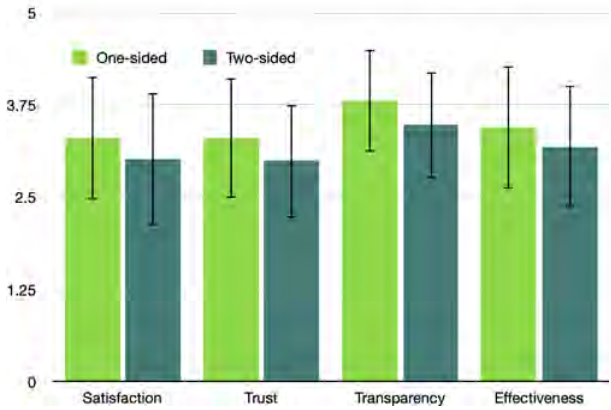

**Figure 2: Average participant ratings of satisfaction, trust, transparency, and effectiveness for one-sided vs. two-sided explanations.**

| One-sided Explanations | | | | |
|---|---|---|---|---|
| | Satisfact. | Trust | Transpar. | Effective. |
| Satisfaction | 1.000 | .509* | .577* | .697* |
| Trust | .509* | 1.000 | .369* | .474* |
| Transparency | .577* | .369* | 1.000 | .658* |
| Effectiveness | .697* | .474* | .658* | 1.000 |
| **Two-sided Explanations** | | | | |
| | Satisfact. | Trust | Transpar. | Effective. |
| Satisfaction | 1.000 | .475* | .443* | .569* |
| Trust | .475* | 1.000 | .398* | .402* |
| Transparency | .443* | .398* | 1.000 | .595* |
| Effectiveness | .569* | .402* | .595* | 1.000 |

* $p < .05$

**Table 3: Correlations between perceptual measures (satisfaction, trust, transparency, effectiveness) for one-sided and two-sided explanations. Higher values in the one-sided condition indicate more internally consistent user experiences.**

$p < .05$). These results suggest that when explanations were one-sided, participants' perceptions of the system were more coherent and internally consistent.

Analysis of participants' open-ended responses provided further insight into their preferences and reactions. A large majority (78.3%) stated that understanding why movies are recommended is important to them. Many participants expressed a desire for richer contextual information in the explanations. Genre emerged as a common theme, with one participant commenting, "The genre of the film, don't want to be surprised when watching what I thought was a comedy turn into a horror" (P6). Others wanted brief summaries or genre labels to accompany recommendations, as in: "A rough summary of the plot and approximately what genre(s) it falls under" (P3). Some also expressed interest in knowing why others appreciated a movie, such as: "Information about why people like it in the first place" (P14). These comments indicate that current explanations may have been too abstract or general for some users.

A particularly notable theme was the placement and impact of negative features in two-sided explanations. Multiple participants reported that ending the explanation with a negative trait left a strong, often unfavorable impression. One participant stated, "Every time you see the explanations end with a negative, instantly I check out and don't really want to watch it" (P6), while another noted that "having the negative aspect at the end leaves a stronger lasting impression" (P58). These responses may help explain the lower ratings for two-sided explanations, as the negative valence—even when minor—was amplified by its position at the end.

To validate the robustness of our findings, we also conducted non-parametric Wilcoxon signed-rank tests. These tests confirmed statistically significant differences between one-sided and two-sided explanations across all four measures (all $p < .05$), with effect sizes ranging from $-0.347$ to $-0.235$. The consistency between parametric and non-parametric analyses reinforces the reliability of our results, while the modest effect sizes suggest that even subtle differences in explanation framing can shape user perceptions.

## 5 DISCUSSION

Our findings suggest a user preference for one-sided explanations in the context of movie recommendations. This challenges some earlier assumptions about the benefits of two-sided explanations, particularly those that emphasize increased transparency or trust through the inclusion of negative aspects [13]. While two-sided approaches may offer advantages in high-stakes or utilitarian decision contexts, our results align more closely with findings from Chen [4], who reported that one-sided messaging can be more persuasive for experience goods such as movies. These outcomes highlight the potential domain-dependence of explanation strategies, suggesting that the effectiveness of an explanation style may be shaped by the nature and stakes of the decision being supported.

The entertainment domain presents specific characteristics that may help explain these results. Selecting a movie on a streaming platform typically involves low commitment and low cost—users can easily abandon a choice if it turns out to be unsatisfactory. This differs from domains such as hotel bookings [9] or product purchases [19], where decisions are less reversible and may carry more risk or cost. In low-stakes environments, users may prefer recommendations that highlight attractive features without drawing attention to potential drawbacks. The stronger correlations we observed between satisfaction and other perception metrics for one-sided explanations suggest that users found these explanations more internally coherent and easier to evaluate positively.

These preferences, however, may not generalize to high-stakes domains such as health, where users often expect more comprehensive and balanced information to support critical decisions. Prior work in health recommender systems has highlighted the need for tailored, context-aware explanation strategies that account for user motivation, behavioral change, and ethical considerations [5]. This contrast underscores the importance of aligning explanation design not only with user preferences but also with the domain and decision complexity involved.

Participants' qualitative feedback reinforces this interpretation. Several noted that the presence of negative information—particularly when placed at the end of an explanation—dampened their interest in the recommended movie. Even when mild, negative features appeared to overshadow the positive aspects, indicating a recency effect or negativity bias. This suggests that the framing and positioning of content within explanations may significantly influence user impressions, even in otherwise well-aligned recommendations. While two-sided explanations might improve transparency by acknowledging trade-offs, in this context, such transparency may have the unintended consequence of reducing user engagement.

Our study has several limitations. First, participants evaluated static explanations in a simulated setting rather than interacting with a functional recommender system. This limits ecological validity and may not fully capture the nuances of real-world decision-making. Second, we used well-known, popular movies to increase familiarity, but this may have introduced bias—participants may have already formed opinions that influenced their responses. Using fictional or lesser-known items in future studies could help control for this. Third, all negative features in our two-sided explanations appeared last, following a "However" clause. This consistent placement may have amplified their impact and should be varied in future experiments to assess order effects. Finally, our sample skewed toward younger participants (84% aged 18–34), which may limit generalizability to broader populations.

Despite these limitations, the findings offer actionable insights for the design of recommender system explanations in entertainment platforms. Our results suggest that emphasizing positive, relevant item attributes may better align with user expectations and preferences in low-stakes contexts. While transparency remains an important design principle, its implementation should consider when and how negative information is presented. Future research could examine how explanation valence preferences vary across user demographics, content genres, or system types. Studies involving interactive interfaces or adaptive explanation strategies may provide more nuanced insights into how users interpret and respond to explanations in real-time decision scenarios.

An additional consideration is the tension between explanation realism and user expectations. While one-sided explanations may feel more appealing in entertainment contexts, they also simplify the nature of recommendation reasoning. Real-world recommendation models often involve trade-offs and conflicting factors, which two-sided explanations may better reflect. However, our results suggest that users may not value this realism when the recommendation task feels subjective or inconsequential. This highlights a potential gap between what is technically accurate and what is perceived as useful or trustworthy by users—especially in hedonic domains. Future work might explore how explanation fidelity influences trust when users are made aware of the underlying algorithmic complexity.

## 6 CONCLUSIONS

This study explored how users perceive one-sided versus two-sided explanations in movie recommendations, addressing three research questions related to perceived quality, the role of negative features, and the interrelationships between perception dimensions. We found that one-sided explanations were rated more positively across all measured aspects—trust, transparency, effectiveness, and satisfaction—and that negative features, when included, appeared to lower user satisfaction. Furthermore, correlations between satisfaction and other perceptual measures were stronger for one-sided explanations, suggesting greater coherence in users' experience of these explanations.

These results suggest that explanation preferences may be shaped by domain-specific factors. In entertainment contexts—where decisions are low-risk and often hedonic—users may favor simpler, positively framed explanations over more balanced or critical ones. While two-sided explanations are often associated with improved transparency, our findings indicate that they may be less well-received in contexts where users expect lightweight, affirming decision support.

These insights have practical implications for the design of recommender systems in streaming services and similar platforms. Focusing on positive, user-aligned features may lead to better engagement and perceived usefulness. At the same time, our results raise broader questions about how explanation effectiveness varies across domains, user types, and content formats.

Future research should investigate how preferences for explanation valence change in interactive settings, across genres, or with personalized vs. generic recommendations. Understanding when and how to introduce critical information without undermining user satisfaction remains an open challenge in explanation design.

More broadly, our findings contribute to ongoing efforts to align explanation strategies with user needs, particularly in domains where engagement and usability outweigh algorithmic transparency. While this study focused on movies, the underlying dynamics of explanation valence and user satisfaction may apply to other domains involving hedonic or low-risk decisions, such as music, casual games, or news personalization. Incorporating explanation design as a user-centric layer in recommender system architecture could support more adaptive and context-aware interfaces. This perspective encourages a shift from "one-size-fits-all" explanation design toward personalization not only of recommendations, but of their justifications.